\documentclass[10pt,conference]{IEEEtran}

\usepackage{listings}
\usepackage{xcolor}

\def \rtpplfontsize {\footnotesize}

\def \rtpplfont {\fontfamily{pcr}\selectfont\rtpplfontsize}

\definecolor{ckeywords}{rgb}{0.13,0.13,1}
\definecolor{ccomments}{rgb}{0,0.5,0.5}
\definecolor{cstrings}{rgb}{0,0.5,0}
\definecolor{cwarnings}{rgb}{1,0.5,0}

\lstdefinelanguage{RTPPL}{
    morekeywords={sensor, actuator,input, output, var, in, if, for, while, update, return, task, sample, observe, infer, delay, read, write, priority, to, offset, importance,
    rate,periodic},
    otherkeywords={->,<-},
    keywordstyle=\color{ckeywords},
    morekeywords=[2]{system, def, model, template},
    keywordstyle=[2]\color{cwarnings},
    morecomment=[l][\color{ccomments}]{//},
    morestring=[b]",
    stringstyle=\color{cstrings},
    sensitive=true,
    basicstyle=\rtpplfont,
    breaklines=true,
    escapeinside={///}{\^^M},
    numbers=left,
    stepnumber=1,
    numberstyle=\tiny,
    xleftmargin=15pt, 
    columns=fixed, 
    showstringspaces=false,
    mathescape=true,
    breaklines=true,
    breakatwhitespace=true,
    mathescape=true,
    showstringspaces=false,
    literate={~}{$\sim$}{1}
}

\lstnewenvironment{rtppl}
  {
    \lstset{
      language=RTPPL,
      numbers=none,
    }
  }{}

\lstnewenvironment{rtppl-lines}
  {
    \lstset{
      language=RTPPL,
    }
  }{}

\newcommand{\rtpplinline}[1]{\lstinline[{language=RTPPL}]|#1|}

\usepackage{listings}
\usepackage{xcolor}

\definecolor{ckeywords}{rgb}{0.13,0.13,1}
\definecolor{ccomments}{rgb}{0,0.5,0.5}
\definecolor{cstrings}{rgb}{0,0.5,0}
\definecolor{cwarnings}{rgb}{1,0.5,0}

\lstdefinelanguage{CorePPL}{
    morekeywords={assume, observe, let, in},
    otherkeywords={},
    keywordstyle=\color{ckeywords},
    morekeywords=[2]{Beta, Bernoulli},
    keywordstyle=[2]\color{cwarnings},
    morecomment=[l][\color{ccomments}]{//},
    morestring=[b]",
    stringstyle=\color{cstrings},
    sensitive=true,
    basicstyle=\rtpplfont,
    breaklines=true,
    escapeinside={---}{\^^M},
    numbers=left,
    stepnumber=1,
    numberstyle=\tiny,
    xleftmargin=15pt, 
    columns=fixed, 
    showstringspaces=false,
    mathescape=true,
    breaklines=true,
    breakatwhitespace=true,
    mathescape=true,
    showstringspaces=false,
    literate={~}{$\sim$}{1}
}

\usepackage{cite}
\usepackage{amsmath,amssymb,amsfonts}
\usepackage{algorithm}
\usepackage[noend]{algpseudocode}
\usepackage{graphicx}
\usepackage{textcomp}
\usepackage{xcolor}
\usepackage{hyperref}
\usepackage{caption,subcaption}

\usepackage{amsthm}
\usepackage{listing}
\usepackage{multicol}
\usepackage{standalone}
\usepackage{siunitx}
\usepackage{etoolbox}
\usepackage{tabularx}
\algblockdefx[Match]{Match}{MatchEnd}
  [2]{\textbf{match} #1 \textbf{with} #2 \textbf{then}}
  [1]{\textbf{else} #1}
\algcblockdefx[Match]{Match}{MatchEls}{MatchEnd}
  [2]{\textbf{else match} #1 \textbf{with} #2 \textbf{then}}
  [1]{\textbf{else} #1}

\begin{document}

\title{Real-Time Probabilistic Programming\\
}

\newtoggle{anonymous}
\togglefalse{anonymous}
\iftoggle{anonymous}{
    \author{\IEEEauthorblockN{Anonymous Author(s)}}
}{
    \author{
    \IEEEauthorblockN{Lars Hummelgren, Matthias Becker, and David Broman}
    \IEEEauthorblockA{EECS and Digital Futures, KTH Royal Institute of Technology, Sweden \\
    \{larshum, mabecker, dbro\}@kth.se}
    }
}

\maketitle

\begin{abstract}
Complex cyber-physical systems interact in real-time and must consider both timing and uncertainty. Developing software for such systems is expensive and difficult, especially when modeling, inference, and real-time behavior must be developed from scratch. Recently, a new kind of language has emerged---called probabilistic programming languages (PPLs)---that simplify modeling and inference by separating the concerns between probabilistic modeling and inference algorithm implementation. However, these languages have primarily been designed for offline problems, not online real-time systems. In this paper, we combine PPLs and real-time programming primitives by introducing the concept of real-time probabilistic programming languages (RTPPL). We develop an RTPPL called ProbTime and demonstrate its usability on an automotive testbed performing indoor positioning and braking. Moreover, we study fundamental properties and design alternatives for runtime behavior, including a new fairness-guided approach that automatically optimizes the accuracy of a ProbTime system under schedulability constraints.
\end{abstract}



\section{Introduction}
\label{sec:intro}

\noindent Probabilistic programming~\cite{GordonEtAl:2014,VanMeentEtAl:2018} is an emerging programming paradigm that enables simple and expressive probabilistic modeling of complex systems. Specifically, the key motivation for \emph{probabilistic programming languages (PPLs)} is to separate the concerns between the model (the probabilistic program) and the inference algorithm. Such separation enables the system designer to focus on the modeling problem without the need to have deep knowledge of Bayesian inference algorithms, such as Sequential Monte Carlo (SMC)~\cite{DoucetEtAl:2001,LundenEtAl:2021} or Markov chain Monte Carlo (MCMC)~\cite{Brooks:1998} methods.

There are many research PPLs, including Pyro~\cite{BinghamEtAl:2019}, WebPPL~\cite{GoodmanStuhlmuller:WebPPL}, Stan~\cite{CarpenterEtAl:2017}, Anglican~\cite{TolpinEtAl:2016}, Turing~\cite{GeXuGhahramani:2018}, Gen~\cite{CusumanoTownerEtAl:2019}, and Miking CorePPL~\cite{LundenEtAl:2022:RootPPL}. These PPLs focus on offline inference problems, such as data cleaning~\cite{LewEtAl:2021}, phylogenetics~\cite{RonquistEtAl:2021}, computer vision~\cite{GothoskarEtAl:2021}, or cognitive science~\cite{GoodmanTenenbaum:2016}, where time and timing properties are not explicitly part of the programming model. Likewise, many languages and environments exist for programming real-time systems~\cite{NatarjanBroman:2018,LohstrohEtAl:2021,BurnsWellings:2007} with no built-in support for probabilistic modeling and inference. Although some recent work exists on using probabilistic programming for reactive synchronous systems~\cite{BaudartEtAl:2020,baudart2022jax} and for continuous time systems~\cite{pfeffer2009ctppl}, no existing work combines probabilistic programming with real-time programming, where both inference and timing constructs are first-class.

Combining probabilistic programming with real-time programming results in several significant research challenges. 
%
Specifically, we identify two key challenges: (i) how to incorporate language constructs for both timing and probabilistic reasoning in a sound manner, and (ii) how to fairly distribute resources among systems of tasks while considering real-time constraints and inference accuracy requirements.


In this paper, we introduce a new kind of language that we call \emph{real-time probabilistic programming languages (RTPPLs)}. In this paradigm, users can focus on the modeling aspect of inferring unknown parameters of the application's problem space
without knowing details of how timing aspects or inference algorithms are implemented. We motivate the ideas behind this new kind of language and outline the main design challenges. To demonstrate the concepts of RTPPL, we develop an RTPPL called ProbTime, which includes probabilistic programming primitives, timing primitives, and primitives for designing modular task-based real-time systems. We create a compiler toolchain for ProbTime and demonstrate how it can be efficiently implemented in an automotive positioning and braking case study. A key aspect of our design is real-time inference using Sequential Monte-Carlo (SMC) and the automatic configuration built on top of this. We have implemented the compiler within the Miking framework~\cite{Broman:2019} as part of the Miking CorePPL effort~\cite{LundenEtAl:2022:RootPPL}.

\begin{figure*}[t!]
\centering
\begin{subfigure}{0.45\textwidth}
\begin{lstlisting}[
    language=CorePPL,
    linerange={2-5}
  ]
let x = assume (Beta 2.0 2.0) in---\label{lst:ppl-code:assume}
observe true (Bernoulli x);---\label{lst:ppl-code:observe}
--- ...\label{lst:ppl-code:dots}
x---\label{lst:ppl-code:result}
\end{lstlisting}
  \caption{Line~\ref{lst:ppl-code:assume} defines the latent variable $\texttt{x}$ and gives the prior $\texttt{Beta(2,2)}$. Line~\ref{lst:ppl-code:observe} shows an observation statement (observing a $\texttt{true}$ value), and line~\ref{lst:ppl-code:result} states that the posterior for $\texttt{x}$ should be computed.}
  \label{fig:ppl-code}
\end{subfigure}
~
\begin{subfigure}{0.15\textwidth}
  \centering
  \includegraphics[width=.9\columnwidth]{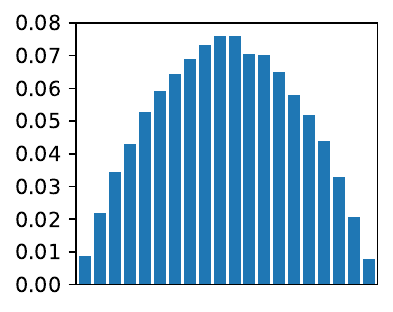}
  \caption{Prior distribution $\texttt{Beta(2,2)}$ for latent variable $\texttt{x}$.}
  \label{fig:coin0}
\end{subfigure}
~
\begin{subfigure}{0.15\textwidth}
  \centering
  \includegraphics[width=.9\columnwidth]{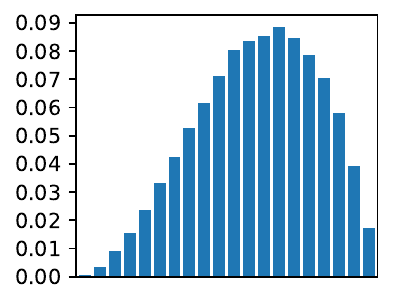}
  \caption{Posterior distribution for $\texttt{x}$, one $\texttt{true}$ observation.}
  \label{fig:coin1}
\end{subfigure}
~
\begin{subfigure}{0.15\textwidth}
  \centering
  \includegraphics[width=.9\columnwidth]{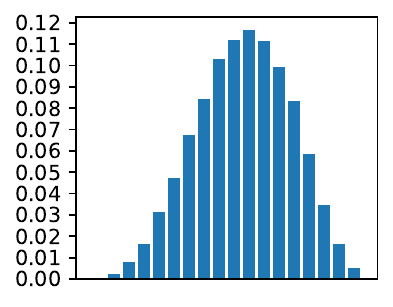}
  \caption{Posterior distri-bution, observations $\texttt{[T, F, F, T, T]}$ }
  \label{fig:coin5}
\end{subfigure}
\caption{A simple CorePPL program modeling a coin flip. Fig.~(\subref{fig:ppl-code}) shows the source code, and Fig.~(\subref{fig:coin0}) the prior distribution. Fig.~(\subref{fig:coin1}) gives the posterior distribution after one coin flip, and Fig.~(\subref{fig:coin5}) the posterior after a sequence of 5 coin flips.}
\vspace{-1em}
\end{figure*}

In summary, we make the following contributions:

\begin{itemize}
\item We introduce the new paradigm of \emph{real-time probabilistic programming languages (RTPPL)}, motivate why it is essential, and outline significant design challenges (Sec.~\ref{sec:motivation}).
\item We design and implement \emph{ProbTime}, a language within the domain of RTPPL (Sec.~\ref{sec:rtppl}), and specify its formal behavior (Sec.~\ref{sec:formal-spec}).
\item We present a novel automated offline configuration approach that maximizes inference accuracy under timing constraints. Specifically, we introduce the concepts of \emph{execution-time} and \emph{particle fairness} within the context of real-time probabilistic programming (Sec.~\ref{sec:system-model-fairness}) and study how they can be incorporated in an automated configuration setting (Sec.~\ref{sec:configuration}).
\item We develop an automotive testbed including both hardware and software. As part of our case study, we demonstrate how a non-trivial ProbTime program can perform indoor positioning and braking in a real-time physical environment (Sec.~\ref{sec:case-study}).
\end{itemize}

\section{Motivation and Challenges with RTPPL}
\label{sec:motivation}

\noindent
This section introduces the main ideas of probabilistic programming, motivates why it is useful for real-time systems, and outlines some of the key challenges. 

\subsection{Probabilistic Programming Languages (PPLs)}
\label{subsec:ppls}

\noindent Probabilistic programming is a rather recent programming paradigm that enables probabilistic modeling where the models can be described as Turing complete programs. Specifically, given a probabilistic program $f$, observed $\textbf{data}$, and \emph{prior} knowledge about the distribution of latent variables $\theta$, a PPL execution environment can automatically approximate the \emph{posterior} distribution of $\theta$. Recall Bayes' rule
\begin{align}
  p(\theta, \textbf{data}) = \frac{p(\textbf{data}, \theta) p(\theta)}{p(\textbf{data})}
  \label{eq:bayes}
\end{align}
\noindent where $p(\theta, \textbf{data})$ is the posterior, $p(\textbf{data}, \theta)$ the likelihood,  $p(\theta)$ the prior, and $p(\textbf{data})$ the normalizing constant. Using a simple toy probabilistic program, we explain how standard PPL constructs relate to Bayes' rule.

Consider Fig.~\ref{fig:ppl-code}, written in CorePPL~\cite{LundenEtAl:2022:RootPPL}, the core language that our work is based on. We use this coin-flip example to give an intuition of the fundamental constructs in a PPL. The model starts by defining a random variable \verb|x| (line~\ref{lst:ppl-code:assume}) using the \verb|assume| construct. In this example, random variable \verb|x| models the probability that a coin flip becomes \verb|true| (we let \verb|true| mean heads and \verb|false| mean tails). For instance, if $p(\texttt{x}) = 0.5$, the coin is fair, whereas if, e.g., $p(\texttt{x}) = 0.7$, it is unfair and more likely to result in \verb|true| when flipped.

The \verb|assume| construct on line~\ref{lst:ppl-code:assume} defines the random variable and gives its prior; in this case, the \verb|Beta| distribution with both parameters equal to $2$. If we sample from this prior, we get the \verb|Beta| distribution depicted in Fig.~\ref{fig:coin0}. Note how the sample constructs in a probabilistic program (\verb|assume| in CorePPL) correspond to the prior $p(\theta)$ in Bayes' rule (Equation~\ref{eq:bayes}).

However, the goal of Bayesian inference is to estimate the \emph{posterior} for a latent variable, given some observations. Line~\ref{lst:ppl-code:observe} in the program shows an \verb|observe| statement, where a coin flip of value \verb|true| is observed according to the Bernoulli distribution. Note how the Bernoulli distribution's parameter depends on the random variable \verb|x|. Consider Fig.~\ref{fig:coin1}, which depicts the inferred posterior distribution given one observed \verb|true| coin flip. As expected, the distribution shifts slightly to the right, meaning that given one sample, the coin is estimated to be somewhat biased toward \verb|true|. Note also how \verb|observe| statements in a probabilistic program correspond to the likelihood $p(\textbf{data}, \theta)$ in Bayes' rule. That is, the likelihood we observe specific \textbf{data}, given a certain latent variable $\theta$.

If we extend Fig.~\ref{fig:ppl-code} with in total five observations \verb|[true, false, false, true, true]| in a sequence (illustrated with \verb|...| on line~\ref{lst:ppl-code:dots}), the resulting posterior looks like Fig.~\ref{fig:coin5}. We can note the following: (i) the resulting mean is still slightly biased toward \verb|true| (we have observed one more \verb|true| value), and (ii) the variance is lower (the peak is slightly thinner). The latter case is a direct consequence of Bayesian inference and one of its key benefits: given the prior and the observed data, the posterior correctly represents the \emph{uncertainty} of the result, not just a point estimate.

There are many inference strategies that can be used for the inference in the toy example, such as sequential Monte Carlo (SMC) or Markov chain Monte Carlo (MCMC). In this work, we focus on SMC and one of its instances, also known as the \emph{particle filter}~\cite{GustafssonEtAl:2002} algorithm, because of its strength in this kind of application. The intuition of particle filters is that inference is done by executing the probabilistic program (the model) multiple times (potentially in parallel). Each such execution point is called a \emph{particle} and consists at the end of a \emph{weight} (how likely the particle is) and the output values, where the set of all particles forms the posterior distribution. Intuitively, the more particles, the better the accuracy of the approximate inference. Many more details of the algorithm (e.g., resampling strategies) are outside this paper's scope, and an interested reader is referred to this technical overview by Naesseth et al.~\cite{naesseth2019elements}. Although hiding the details of the inference algorithm is one of the key ideas of PPLs (separating the model from inference), the \emph{number of particles} is essential for automatic inference and scheduling. Hence, this is a key concept in our work and will be discussed further in this paper.

\subsection{Real-Time Probabilistic Programming (RTPPL)}
\label{subsec:rt+ppl}

\begin{figure*}[t!]
\centering
\begin{subfigure}{0.45\textwidth}
\begin{lstlisting}[
    language=RTPPL,
    linerange={46-61}
  ]
system {
  sensor speed1 : Float rate 50ms///\label{lst:model:sa1}
  sensor speed2 : Float rate 100ms
  actuator brake : Float rate 500ms///\label{lst:model:sa2}
  task speedEst1 = Speed(250ms) importance 1///\label{lst:model:tasks1}
  task speedEst2 = Speed(500ms) importance 1
  task pos = Position() importance 3
  task braking = Brake(300ms) importance 4///\label{lst:model:tasks2}
  speed1 -> speedEst1.in1///\label{lst:model:conn1}
  speed2 -> speedEst2.in1
  speedEst1.out -> pos.in1///\label{lst:model:s1conn1}
  speedEst1.out -> braking.in1///\label{lst:model:s1conn2}
  speedEst2.out -> pos.in2
  speedEst2.out -> braking.in2
  pos.out -> braking.in3
  braking.out -> brake }///\label{lst:model:conn2}
\end{lstlisting}
  \caption{The system declaration of a ProbTime program.}
  \label{fig:model-code}
\end{subfigure}
~
\begin{subfigure}{0.49\textwidth}
  \centering
  \includegraphics[width=\columnwidth]{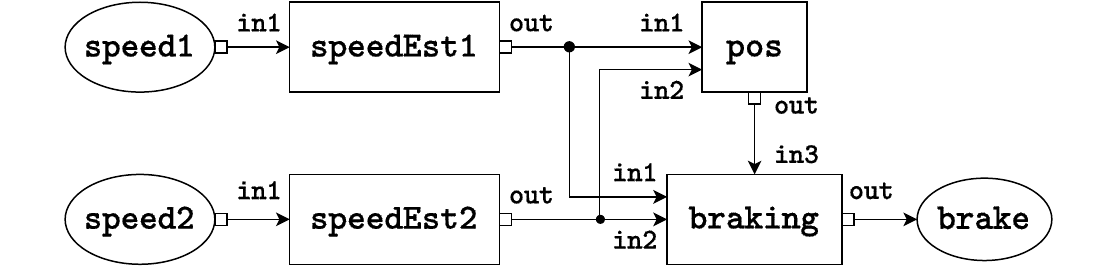}
  \caption{A graphical representation of the system defined in Fig.~\ref{fig:model-code}. The system consists of two sensors, \texttt{speed1} and \texttt{speed2} on the left-hand side, four tasks \texttt{speedEst1}, \texttt{speedEst2}, \texttt{pos}, and \texttt{braking} in the middle, and an actuator \texttt{brake} on the right-hand side. Lines represent connections, where the white box represents the output port and the incoming arrow represents the input port.}
  \label{fig:model-ex}
\end{subfigure}
\caption{Example system of a ProbTime program in textual (\subref{fig:model-code}) and graphical form (\subref{fig:model-ex}).}
\label{fig:probtime-model}
\vspace{-1.5em}
\end{figure*}

\noindent The toy example of a coin flip in the previous example gives an intuition of probabilistic programming but does not show its power in large-scale applications. Probabilistic programming is used in several domains today, but surprisingly, little has been done in the context of time-critical applications. Although a rich literature exists on using Bayesian inference and filtering algorithms (e.g., particle filters in aircraft positioning~\cite{GustafssonEtAl:2002}), these algorithms are typically hand-coded where both the probabilistic model and the inference algorithm are combined. This means that a developer needs to have deep insights into three different fields: (i) probabilistic modeling, (ii) Bayesian inference algorithms, and (iii) real-time programming aspects. Standard probabilistic programming is motivated by the appealing argument of separating (i) and (ii), thus enabling probabilistic modeling by developers without deep knowledge of how to implement inference efficiently and correctly. However, so far, probabilistic programming has not been combined with (iii) real-time aspects, such as timing, scheduling, and worst-case execution time estimation. This paper aims to make the first step towards mitigating this gap. We call this approach \emph{real-time probabilistic
programming (RTPPL)}.

How can an RTPPL be designed? For instance, we can extend an existing PPL with timing semantics, extend an existing real-time language with PPL constructs, or create a new \emph{domain-specific language (DSL)} including only a minimal number of constructs for reasoning about timing and probabilistic inference. In this research work, we choose the latter to avoid complexity when extending large existing languages. In particular, our purpose is to create this rather small research language to study the \emph{fundamentals} of this new kind of language category. We identify and examine the following key challenges with RTPPLs:
\begin{itemize}
\item \textbf{Challenge A:} In an RTPPL, how can we encode real-time properties (deadlines, periodicity, etc.) and probabilistic constructs (sampling, observations, etc.) without forcing the user to specify low-level details such as particle counts and worst-case execution times?
\item \textbf{Challenge B:} How can a runtime environment be constructed, such that a \emph{fair} amount of time is spent on different tasks, giving the right number of particles for each task (for accuracy), where the system is still known to be statically schedulable?
\end{itemize}

\noindent We first address Challenge A (Sec.~\ref{sec:rtppl}-\ref{sec:formal-spec}) by designing a new minimalistic research DSL called ProbTime and defining its formal behavior. Using the new DSL, we address Challenge B by studying two new concepts: \emph{execution-time fairness} and \emph{particle fairness} (Sec.~\ref{sec:system-model-fairness}), and how particle counts, execution-time measurements, and scheduling can be performed automatically to get a fair and working configuration (Sec.~\ref{sec:configuration}).

\section{ProbTime - an RTPPL}
\label{sec:rtppl}

\noindent In this section, we introduce a new real-time probabilistic programming language called ProbTime.
We present the timing and probabilistic constructs in the language using a small example, followed by an overview of the compiler implementation. ProbTime is open-source and publically available\footnote{\url{https://github.com/miking-lang/ProbTime}}.

\subsection{System Declaration}
\label{subsec:system-overview}

\noindent We present a ProbTime system declaration implementing automated braking for a train in Fig.~\ref{fig:probtime-model}. The system keeps track of the train's position and uses the automatic brakes before reaching the end of the track. The train provides observation data via two speed sensors with varying frequencies and accuracy (left side of Fig.~\ref{fig:model-ex}), and the system can activate the brakes via an actuator (right side of Fig.~\ref{fig:model-ex}). In Fig.~\ref{fig:model-code}, we first declare the sensors and actuators and associate them with a type to indicate what kind of data they operate on (lines~\ref{lst:model:sa1}-\ref{lst:model:sa2}). We also specify the rate at which data is provided by sensors and consumed by actuators. The squares in the graphical representation correspond to the tasks of our system. We instantiate the tasks on lines~\ref{lst:model:tasks1}-\ref{lst:model:tasks2}. Note that \texttt{speedEst1} and \texttt{speedEst2} are instantiated from the same template.

The \texttt{importance} construct captures the relative importance of tasks. We use the importance value to indicate how important the quality of inference performed by a task is. It is not to be confused with the task priority.
%
%
As we saw in the coin flip example (Sec.~\ref{sec:motivation}), the number of particles used in inference relates to the quality of the estimate. In the train example, we use \verb|importance| on lines~\ref{lst:model:tasks1}-\ref{lst:model:tasks2} to indicate that \texttt{pos} and \texttt{braking} should run more particles than \texttt{speedEst1} and \texttt{speedEst2}. We discuss importance in more detail in Sec.~\ref{sec:system-model-fairness}.

On lines~\ref{lst:model:conn1}-\ref{lst:model:conn2} in Fig.~\ref{fig:model-code}, we declare the connections using arrow syntax \verb|a -> b|, specifying that data written to output port \verb|a| is delivered to input port \verb|b|. Names of sensors and actuators represent ports and \verb|x.y| refers to port \verb|y| of task \verb|x|.

\subsection{Templates and Timing Constructs}
\label{subsec:template-timing}

\noindent Continuing with the train braking example, we outline the definitions of the task templates \texttt{Speed} and \texttt{Position} in Listing~\ref{lst:tasktemplates}. Consider the definition of the \texttt{Speed} template (lines~\ref{lst:model:speed1}-\ref{lst:model:speed2}). We declare the input port \texttt{in1} and the output port \texttt{out} on lines~\ref{lst:model:speedin1}-\ref{lst:model:speedout}, and annotate them with types.

We control timing in ProbTime using the iterative \verb|periodic| block statement. A \verb|periodic| block is a loop where each iteration starts after a statically known delay. Each iteration constitutes a task instance. On lines~\ref{lst:model:speed-periodic}-\ref{lst:model:speed2}, we define a periodic block whose period is determined by the \verb|period| argument of the \texttt{Speed} template. For instance, if \verb|period| is \SI{1}{\second}, the release time is one second after the release time of the previous iteration. Our approach to timing is similar to the logical execution time paradigm~\cite{DBLP:conf/birthday/KirschS12}. However, while all timestamps are absolute under the hood, we only expose a relative view of the logical time of the current task instance.

The \verb|read| statement retrieves a sequence of inputs available from an input port at the logical time of the task instance. On line~\ref{lst:model:speed-read}, we retrieve inputs from port \verb|in1| and store them in the variable \verb|obs|. Similarly, the \verb|write| statement (line~\ref{lst:model:speed-write}) sends a message with a given payload to the specified output port.

In \texttt{Position} (lines~\ref{lst:model:pos1}-\ref{lst:model:pos2}), we use our prior belief of the train's position when estimating its current position. However, variables in ProbTime are immutable by default. We use \verb|update d| in the periodic block on line~\ref{lst:model:pos-upd} to indicate that updates to \verb|d| (the position distribution) should be visible in subsequent iterations. Also, in the \verb|write| on line~\ref{lst:model:pos-write}, we specify an offset to the timestamp of the message relative to the current logical time of the task (it is zero by default).

\begin{lstlisting}[
  language=RTPPL,
  linerange={20-36},
  caption={Definition of the \texttt{Speed} and \texttt{Position} task templates.},
  label={lst:tasktemplates},
  float=t!
]
template Speed(period : Int) {///\label{lst:model:speed1}
  input in1 : Float///\label{lst:model:speedin1}
  output out : Dist(Float)///\label{lst:model:speedout}
  periodic period {///\label{lst:model:speed-periodic}
    read in1 to obs///\label{lst:model:speed-read}
    infer speedModel(obs) to d///\label{lst:model:speed-infer}
    write d to out } }///\label{lst:model:speed-write} \label{lst:model:speed2}
template Position() {///\label{lst:model:pos1}
  input in1 : Dist(Float)
  input in2 : Dist(Float)
  output out : Dist(Float)
  infer initPos() to d
  periodic 1s update d {///\label{lst:model:pos-upd}
    read in1 to speed1
    read in2 to speed2
    infer positionModel(d, speed1, speed2) to d
    write d to out offset 500ms } }///\label{lst:model:pos-write} \label{lst:model:pos2}
\end{lstlisting}

\subsection{Models and Probabilistic Constructs}
\label{subsec:model-prob}

\noindent We define three probabilistic constructs in ProbTime, similar to what is used in other PPLs: \texttt{sample}, \texttt{observe}, and \texttt{infer}. We use \texttt{sample} and \texttt{observe} when defining probabilistic models, and we use the \texttt{infer} construct to produce a distribution from a probabilistic model in task templates.

\begin{lstlisting}[
  language=RTPPL,
  linerange={10-19},
  caption={Probabilistic model for estimating the train's speed.},
  label={lst:train-speed-model},
  float=t!,
  belowskip=-1em
]
model speedModel(obs : [TSV(Float)]) : Float {///\label{lst:model:model1}
  sample b ~ Uniform(0.0, maxSpeed)///\label{lst:model:intercept}
  sample m ~ Gaussian(0.0, 1.0)
  sample sigma ~ Gamma(1.0, 1.0)///\label{lst:model:sigma}
  for o in obs {///\label{lst:model:for1}
    var ts = timestamp(o)///\label{lst:model:to-timestamp}
    var x = intToFloat(ts) / intToFloat(1s)///\label{lst:model:tstofloat}
    var y = m * x + b
    observe value(o) ~ Gaussian(y, sigma) }///\label{lst:model:observe} \label{lst:model:for2}
  return b }///\label{lst:model:model-ret} \label{lst:model:model2}
\end{lstlisting}

Consider the probabilistic model defined using the function \texttt{speedModel} in Listing~\ref{lst:train-speed-model}. The function models the speed of the train at the current logical time using Bayesian linear regression. The parameter \texttt{obs} is a sequence of speed observations encoded as a sequence of floating-point messages (we use \texttt{TSV} as a short-hand for timestamped values).

In this model, we use a \texttt{sample} statement to sample a random slope and intercept of a line, assuming acceleration is constant. There are three latent variables (defined on lines~\ref{lst:model:intercept}-\ref{lst:model:sigma}): \verb|b| (the offset of the line), \verb|m| (the slope of the line), and \verb|sigma| (the variance, approximating the uncertainty of the estimate). The estimated line can be seen as a function of the speed over (relative) time, where $0$ is the logical time of the task instance.

\begin{figure}[t!]
\centering
\includegraphics[width=.8\columnwidth]{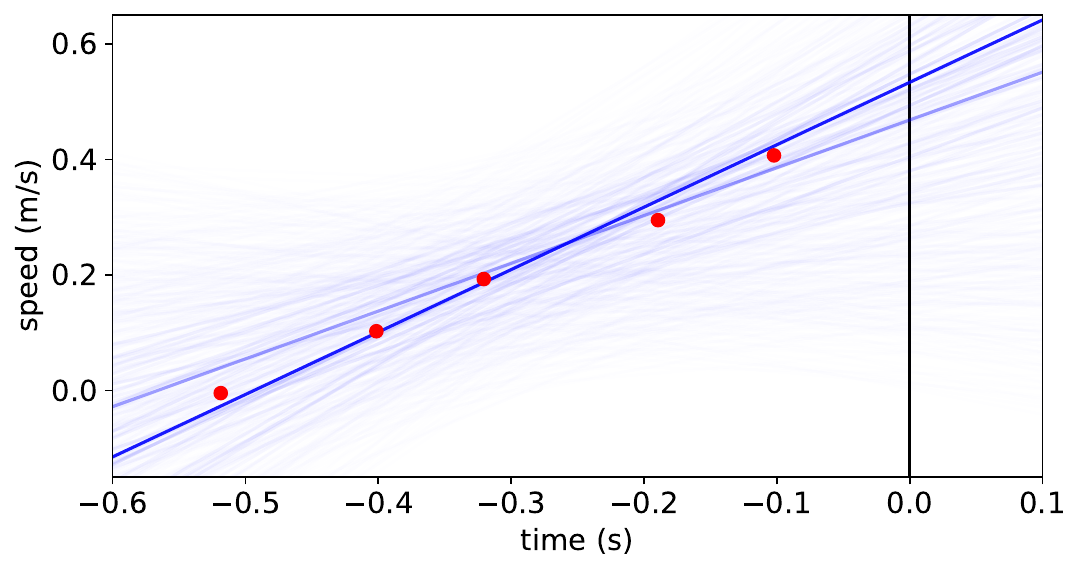}
\caption{Outcome of the speed model for the \texttt{speedEst2} task on synthetic data. The x-axis represents time in seconds, and the y-axis represents the train's speed in meters per second (we seek the speed at the current time, $x = 0.0$). The red dots represent speed observations, and the blue lines are estimates from the model (slope and intercept), where the opacity of a line indicates its likelihood.}
\label{fig:speed-line}
\vspace{-1em}
\end{figure}

In the for-loop on lines~\ref{lst:model:for1}-\ref{lst:model:for2}, we update our belief based on the speed observations. First, we translate the relative timestamp of the observation \verb|o| to a floating-point number, representing our x-value (lines~\ref{lst:model:to-timestamp}-\ref{lst:model:tstofloat}). We use the \texttt{observe} statement on line~\ref{lst:model:observe} to update our belief. The observed value is the data of \verb|o| (retrieved using \verb|value(o)|), and we expect observations to be distributed according to a Gaussian distribution centered around the y-value. We return the train's estimated speed (the intercept of the line) on line~\ref{lst:model:model-ret}. Figure~\ref{fig:speed-line} presents a sample plot of what this looks like when the train accelerates. As we use relative timestamps in ProbTime, we can define this rather complicated model succinctly.

\begin{figure*}[t!]
\centering
\includegraphics[width=.9\textwidth]{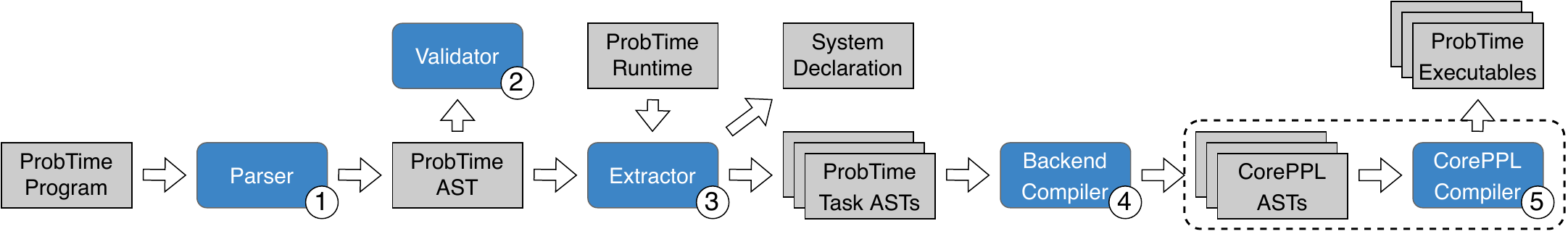}
\caption{An overview of the ProbTime compiler. Gray square boxes represent artifacts, and rounded blue squares represent compiler transformations (numbered by the order in which they run). We use the pre-existing CorePPL compiler to produce executable code in the dashed area.}
\label{fig:rtppl-compiler}
\vspace{-1.5em}
\end{figure*}

Finally, recall Listing~\ref{lst:tasktemplates} at line \ref{lst:model:speed-infer}, where we use the \texttt{infer} statement to infer the posterior using the probabilistic model (in this case \texttt{speedModel}).

\subsection{Compiler Implementation}
\label{subsec:compiler-implementation}

\noindent Fig.~\ref{fig:rtppl-compiler} depicts the key steps of the ProbTime compiler. The compiler is implemented within the Miking framework~\cite{Broman:2019} because of its good support for defining domain-specific languages and the possibility of extending the Miking CorePPL~\cite{LundenEtAl:2022:RootPPL} framework with real-time constructs.

A key step of the compilation is the extractor (3 in Fig.~\ref{fig:rtppl-compiler}), which produces one abstract syntax tree (AST) for each task declared in the ProbTime AST. We combine this AST with a runtime AST, which, for instance, implements \verb|infer| and \verb|delay|. We individually pass the task ASTs to the backend compiler (4), translating them to CorePPL ASTs, which are compiled into executables using the existing CorePPL compiler (5). Our implementation consists of 4400 lines of code.

\section{System Model and Formal Specification}
\label{sec:formal-spec}

This section presents a formal specification of ProbTime on a system level (Sec.~\ref{subsec:formal-system}), including a system model. We also consider ProbTime on a task level (Sec.~\ref{subsec:formal-task}), focusing on communication and the probabilistic constructs.

\subsection{System-Level Specification}
\label{subsec:formal-system}

We define a ProbTime system as a tuple $(\mathcal{T}, \mathcal{S}, \mathcal{A}, \mathcal{C})$, where $\mathcal{T}$ is a set of tasks, $\mathcal{S}$ is a set of sensors, $\mathcal{A}$ is a set of actuators, and $\mathcal{C}$ is a set of connections. A task $\tau$ is represented by the tuple $(C, T, \pi, n, v)$, where $C$ is its worst-case execution time (WCET), $T$ its period, $\pi$ its static priority, $n$ the number of particles, and $v$ its assigned importance value. Tasks are subject to an implicit deadline $D = T$. We assume the \verb|periodic| block of each task $\tau$ contains at most one use of \verb|infer|, for which we set the number of particles $n_\tau$.

We assume the tasks of $\mathcal{T}$ are ordered by increasing period. The tasks execute on an exclusive subset of the cores available on a platform. Specifically, if the platform has $m$ cores, we schedule our tasks on a subset $\mathcal{U}$ of cores, where $k < m$ (we reserve one core for the platform). Each task $\tau$ is statically assigned a core $c \in \mathcal{U}$. We use partitioned fixed-priority scheduling on the cores $\mathcal{U}$. Priorities are assigned on the rate monotonic principle~\cite{liu1973}, where the task $\tau$ with the lowest period $T_\tau$ has the highest priority $\pi_\tau$. The tasks are preemptive.

Each task $\tau$ has a set of input ports $I_\tau$ from which it can receive data and a set of output ports $O_\tau$ to which it can send data. Sensors act as output ports, and actuators as input ports. A ProbTime system has a set of input ports $I = \mathcal{A} \cup \bigcup_{\tau \in \mathcal{T}} I_\tau$ and a set of output ports $O = \mathcal{S} \cup \bigcup_{\tau \in \mathcal{T}} O_\tau$. The set $\mathcal{C}$ contains connections $c_i = (o_j, i_k), o_j \in O \land i_k \in I$. 
We say that a task $\tau$ is a \emph{predecessor} of $\tau'$ if $\exists (o, i) \in \mathcal{C}. o \in O_{\tau} \land i \in I_{\tau'}$.

Every port $p \in I \cup O$ of a task has an associated buffer $B_p$ storing incoming or outgoing messages. Given the particle counts $n_\tau$ of all tasks $\tau$, we can determine the message size. The compiler counts the number of times a task writes to each port when this can be determined at compile-time (otherwise, the compiler produces a warning). Based on these counts and the declared \verb|rate| of sensors and actuators, the automatic configuration (Sec.~\ref{sec:configuration}) verifies that a given (fixed) buffer size is sufficient for the messages passing through all ports.

\subsection{Task-Level Specification}
\label{subsec:formal-task}

We present the behavior of a task $\tau$ in Algorithm~\ref{alg:task-runtime}. The \textsc{RunTaskInstance} function (lines~\ref{alg:task-runtime:rti1}-\ref{alg:task-runtime:rti2}) is called at the start of the \verb|periodic| block. It starts with an absolute delay until the next period, increasing the logical time~\cite{LohstrohEtAl:2023} of the task by $T_\tau$ (line~\ref{alg:task-runtime:delay}). The task $\tau$ reads all data from input ports $i \in I_\tau$ to buffers $B_i$ (using \textsc{ReadNonBlocking} on line~\ref{alg:task-runtime:read1}), runs an iteration of the \verb|periodic| block (such as lines~\ref{lst:model:speed-read}-\ref{lst:model:speed-write} in Listing~\ref{lst:tasktemplates}) in \textsc{RunIteration} (line~\ref{alg:task-runtime:exec}), and writes messages stored in output buffers $B_o$ to their output ports $o \in O_\tau$ (using \textsc{WriteNonBlocking} on line~\ref{alg:task-runtime:write1}).

Reads and writes in a ProbTime task operate on the buffers rather than directly on the ports. The statement \rtpplinline{read i to x} results in $\texttt{x} \gets \Call{ReadInputPort}{\texttt{i}}$. Similarly, \rtpplinline{write v to o offset d} results in $\Call{WriteOutputPort}{\texttt{v}, \texttt{o}, \texttt{d}}$, which concatenates a message with payload \verb|v| and timestamp equal to the current logical time of the task ($t_0$) plus an offset (\verb|d|) to the buffer $B_o$. The reads from and writes to the ports are non-blocking, so we have no precedence relationships between tasks, and communication has no impact on scheduling.

\begin{figure}[t!]
\vspace{-1em}
\begin{algorithm}[H]
\vspace{-0.25em}
\footnotesize
\caption{Definition of the behavior of the task runtime.}
\label{alg:task-runtime}
\begin{algorithmic}[1]
\Function{RunTaskInstance$_\tau$}{\null} \label{alg:task-runtime:rti1}
\While{true}
\State $\Call{DelayUntil}{T_\tau}$ \label{alg:task-runtime:delay}
\For{$i \in I_\tau$} \label{alg:task-runtime:read1}
    $B_i \gets \Call{ReadNonBlocking}{i}$
\EndFor \label{alg:task-runtime:read2}
\State $\Call{RunIteration}{\tau}$ \label{alg:task-runtime:exec}
\For{$o \in O_\tau$} \label{alg:task-runtime:write1}
    $\Call{WriteNonBlocking}{o, B_o}$
\EndFor \label{alg:task-runtime:write2}
\EndWhile
\EndFunction \label{alg:task-runtime:rti2}
\Function{ReadInputPort}{$i$}
\State \Return $B_i$
\EndFunction
\Function{WriteOutputPort}{$v, o, d$}
\State $B_o \gets B_o, (v, t_0 + d)$
\EndFunction
\end{algorithmic}
\end{algorithm}
\vspace{-2em}
\end{figure}

\begin{figure}[t!]
\begin{algorithm}[H]
\vspace{-0.25em}
\footnotesize
\caption{Algorithmic definition of the formal behavior of probabilistic constructs within a task.}
\label{alg:formal}
\begin{algorithmic}[1]
\Function{Infer$_\tau$}{$m, x_1, \ldots, x_n$} \label{alg:formal:infer1}
\For{$i \in 1 \ldots n_\tau$}
\State $v_i, w_i \gets \Call{RunParticle}{m, x_1, \ldots, x_n}$
\EndFor
\State \Return $(v_1, w_1), \ldots, (v_{n_\tau}, w_{n_\tau})$ \label{alg:formal:infer-ret}
\EndFunction \label{alg:formal:infer2}
\Function{RunParticle}{$m, x_1, \ldots, x_n$} \label{alg:formal:runpart1}
\State $w \gets 0$
\State $v \gets m_w(x_1, \ldots, x_n)$ \label{alg:formal:runpart-call}
\State \Return $v, w$
\EndFunction \label{alg:formal:runpart2}
\Function{UpdateWeight}{$w, o, d$} \label{alg:formal:updatew1}
\State \Return $w + \Call{LogObserve}{o, d}$
\EndFunction \label{alg:formal:updatew2}
\Function{LogObserve}{$o, d$} \label{alg:formal:logobs1}
\If{$\Call{Elementary}{d}$} \Return $\text{logPdf}_d(o)$ \label{alg:formal:logpdf}
\Else\ \textbf{error}
\EndIf
\EndFunction \label{alg:formal:logobs2}
\Function{Sample}{$d$} \label{alg:formal:sample1}
\If{$\Call{Elementary}{d}$} \Return $\text{sample}_d()$ \label{alg:formal:sample-elem}
\Else
    \State $r \gets \Call{SampleUniform}{0, c_{|d|}}$ \label{alg:formal:sample-emp1}
    \State $j \gets \Call{LowerBound}{r, c}$
    \State \Return $v_j$ \label{alg:formal:sample-emp2}
\EndIf
\EndFunction \label{alg:formal:sample2}
\end{algorithmic}
\end{algorithm}
\vspace{-2em}
\end{figure}

We present the formal behavior of the probabilistic constructs \verb|infer|, \verb|observe|, and \verb|sample| in Algorithm~\ref{alg:formal}. Consider the definition of inference for a task $\tau$ in the \textsc{Infer}$_\tau$ function (lines~\ref{alg:formal:infer1}-\ref{alg:formal:infer2}). We perform inference by running a fixed number of particles $n_\tau$, as determined by the automatic configuration (Sec.~\ref{sec:configuration}). Each particle $(v_j, w_j)$ consists of a value $v_j$ returned by the probabilistic model function $m$ and a weight $w_j$ determined by use of \verb|observe| within the model.
The sequence of particles returned by \textsc{Infer}$_\tau$ (line~\ref{alg:formal:infer-ret}) is the resulting distribution.
The statement \rtpplinline{infer m(x, y) to d} corresponds to $\texttt{d} \gets $\textsc{Infer}$_\tau(\texttt{m}, \texttt{x}, \texttt{y})$ in our pseudocode.

When we run a particle in the \textsc{RunParticle} function (lines~\ref{alg:formal:runpart1}-\ref{alg:formal:runpart2}), we use a weight $w$, representing the accumulated log-likelihood of a particle having a particular value $v$. Every statement \rtpplinline{observe o ~ d} modifies the weight as $w 
\gets \Call{UpdateWeight}{w, \texttt{o}, \texttt{d}}$. We denote the model function as $m_w$ on line~\ref{alg:formal:runpart-call} to indicate that the model $m$ mutates the weight $w$. For instance, if the model $m_w$ is \texttt{speedModel} of Listing~\ref{lst:train-speed-model}, we run lines~\ref{lst:model:intercept}-\ref{lst:model:model-ret}, and update $w$ for each use of \rtpplinline{observe} on line~\ref{lst:model:observe}. The \textsc{LogObserve} function (lines~\ref{alg:formal:logobs1}-\ref{alg:formal:logobs2} in Algorithm~\ref{alg:formal}) represents the computation of the logarithmic weight for an \verb|observe| statement. ProbTime supports observations on elementary distributions (e.g., \verb|Gaussian|) but not empirical distributions (produced by an \verb|infer|). For an elementary distribution, we compute the weight using its logarithmic probability distribution function (line~\ref{alg:formal:logpdf}).

When we sample a random value as in \rtpplinline{sample x ~ d}, we assign a random value to the variable \verb|x| from the distribution \verb|d|. We define this in the \textsc{Sample} function (lines~\ref{alg:formal:sample1}-\ref{alg:formal:sample2}).
For an elementary distribution, we use a known sampling function (line~\ref{alg:formal:sample-elem}).
For empirical distributions, we use the cumulative weights $c_j = \sum_{i=1}^j w_i$ to pick a random particle through a binary search (lines~\ref{alg:formal:sample-emp1}-\ref{alg:formal:sample-emp2}). Therefore, the time complexity of sampling is $O(\log n)$ for an empirical distribution of $n$ particles. As ProbTime supports sending empirical distributions between tasks, this complexity results in a dependency. A task $\tau_i$ that samples from a distribution sent by a predecessor $\tau_j$ has an increased execution time if we increase $n_{\tau_j}$.

\section{Fairness}
\label{sec:system-model-fairness}

\noindent In this section, we present new perspectives on fairness that emerge when combining real-time and probabilistic inference.

\subsection{Execution-Time and Particle Fairness}
\label{subsec:fairness}

\noindent We know that more particles improve the accuracy of the inference, but it is not obvious how the result of one task $\tau$ impacts the whole system's behavior. As we concluded in Sec.~\ref{sec:formal-spec}, increasing the execution time or particle count of a task may affect the performance of other tasks. ProbTime allows the user to specify how important each task is. Specifically, we use the importance values $v_\tau$ associated with each task $\tau$ as a measure to guide fairness. But what is the right approach to fairness? We define and study fairness in two ways: (i) \emph{execution-time fairness}, where we allocate execution time budgets $B_\tau$ (which the WCETs $C_\tau$ of each task $\tau$ must not exceed) to tasks proportional to their importance values, and (ii) \emph{particle fairness}, where fairness between tasks is given in terms of the number of particles used in the inference.

\begin{figure}[t!]
\centering
\includegraphics[width=.8\columnwidth]{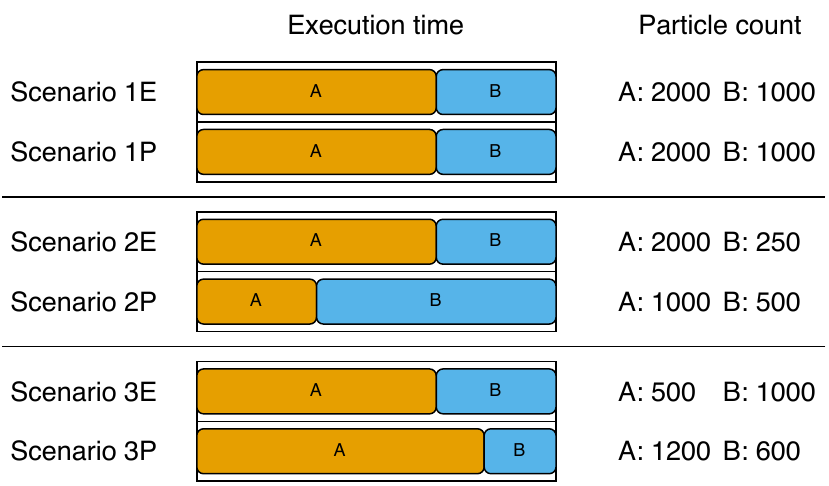}
\caption{Illustration of execution time and particle counts for tasks $\tau_A$ and $\tau_B$ ($\tau_A$ is twice as important as $\tau_B$ and both have the same period) for three scenarios when considering execution-time fairness (E) and particle fairness (P).}
\label{fig:fairness-etpc}
\vspace{-1em}
\end{figure}

Consider Fig.~\ref{fig:fairness-etpc}, where we present three scenarios to exemplify the differences between execution-time and particle fairness. Assume we have two tasks $\tau_A$ and $\tau_B$ such that $T_{\tau_A} = T_{\tau_B}$ and $v_{\tau_A} = 2 \cdot v_{\tau_B}$, running on one core. The tasks are independent and they have the same execution time per particle. We present the outcome of the three scenarios (labeled 1, 2, and 3, respectively) for $\tau_A$ and $\tau_B$ concerning their relative ratios of execution time and particle count.

In scenario 1, we find in both cases that, because the tasks have the same execution time per particle, they both end up with the same ratio of execution time and particles.

In scenario 2, we assume task $\tau_B$ requires four times as much execution time as $\tau_A$ per particle. 
If we use execution-time fairness, the execution time ratio of the tasks remains the same, but $\tau_B$ only has time to produce $250$ particles. For particle fairness, the slowdown of $\tau_B$ results in both tasks running fewer particles to maintain a fair allocation of particle counts, skewing the execution time in favor of $\tau_B$.

In scenario 3, we instead assume $\tau_A$ samples from a distribution sent from $\tau_B$, meaning the execution time of $\tau_A$ depends on the particle count of $\tau_B$. The exact outcome depends on the ratio of time $\tau_A$ spends on sampling. Using execution-time fairness, we find that $\tau_A$ only has time to produce $500$ particles. Task $\tau_B$ produces the same number of particles as in scenario 1 because it is independent of $\tau_A$. In contrast, when using particle fairness, $\tau_A$ produces more particles, and $\tau_B$ produces fewer.

Which kind of fairness is preferred in a real-time probabilistic inference setting?
Although execution-time fairness may seem natural from a real-time perspective, we will see (surprisingly) in the evaluation (Sec.~\ref{subsec:experiments-results}) that particle fairness works better in practice.

Fairness is applicable in scenarios where the output quality is proportional to the time spent producing it. This paper focuses explicitly on SMC inference, where the number of particles controls the quality. However, we foresee no issues extending this to, e.g., MCMC inference algorithms by considering the number of runs instead of the particle count.

\subsection{Computation Maximization}
\label{subsec:utilization}

\noindent So far, we only considered the case where tasks run on one core. What if we want to apply fairness to tasks running on multiple cores? We consider two extremes: prioritizing fairness or prioritizing utilization. If we prioritize fairness, we allocate a fixed amount of resources (execution time or particles) to a task instance proportional to its importance value, regardless of which core the task is mapped to. We refer to this as \emph{fair utilization}. If we prioritize utilization, we maximize the utilization on each core individually, ignoring the importance of tasks running on other cores. We refer to this as \emph{maximum utilization}. In both cases, the resulting system is schedulable.

\begin{figure}[!t]
\centering
\begin{subfigure}[c]{0.4\columnwidth}
\begin{tabular}{clll}
$i$ & $T_i$ & $v_i$ & $c$\\
\hline
$A$ & \SI{1}{\second} & 8 & 1\\
$B$ & \SI{0.5}{\second} & 4 & 1\\
$C$ & \SI{1}{\second} & 4 & 2\\
$D$ & \SI{0.25}{\second} & 2 & 2
\end{tabular}
\end{subfigure}%
\hfill%
\begin{subfigure}[c]{0.59\columnwidth}
\centering
\includegraphics[width=.8\columnwidth]{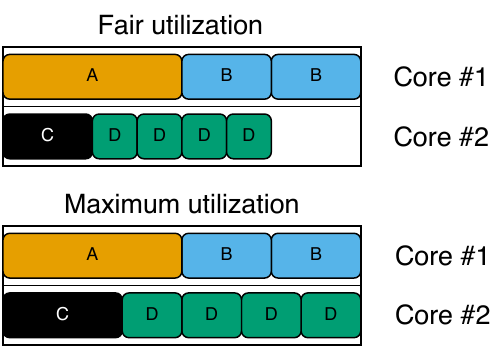}
\end{subfigure}
\caption{The resulting utilization of the tasks (left) allocated on two cores when prioritizing fairness (top right) or maximizing the utilization (bottom right) in terms of execution time.}
\label{fig:fairness-util}
\vspace{-1.5em}
\end{figure}

Assume we allocate the tasks listed on the left-hand side of Fig.~\ref{fig:fairness-util} on a dual-core system (where $T_i$ is the period, $v_i$ is the importance, and $c$ is the core of task $\tau_i$).
On the right-hand side, we present the result of applying the two alternatives using execution-time fairness (this also applies to particle fairness). Each box corresponds to a task instance ($\tau_D$ has four times the frequency of $\tau_A$; hence, it has four boxes), and we consider importance values to apply per task instance. We see that $\tau_A$ and $\tau_B$ are allocated the same execution time, as $\tau_A$ is twice as important as $\tau_B$, but $\tau_B$ runs twice as frequently.


The main takeaway from Fig.~\ref{fig:fairness-util} is that, when prioritizing fair utilization, all cores are not always fully utilized. Recall that the tasks of the system may depend on each other. Increasing the execution times of $\tau_C$ and $\tau_D$ may negatively impact the performance of $\tau_A$ and $\tau_B$. This is unfair, considering that $\tau_A$ and $\tau_B$ have higher importance. On the other hand, if the tasks are independent (which they are in this example), we are wasting computational resources. As an alternative, we also propose the concept of \emph{fair utilization maximization}, where we start from fair utilization and gradually increase the resources (execution time or particles) allocated to tasks on cores that are not fully utilized as long as the system remains schedulable. If $\tau_A$ and $\tau_B$ do not depend on $\tau_C$ and $\tau_D$, we get the same result as when prioritizing maximum utilization.

\section{Automatic Configuration}
\label{sec:configuration}

\noindent In this section, we present two approaches to automatic configuration that we apply to a compiled ProbTime system to maximize inference accuracy while preserving schedulability. Automatic configuration is performed before running the system so it does not introduce runtime overheads.


\subsection{Configuration Overview}
\label{subsec:config-overview}

\noindent We define two key phases needed to configure a ProbTime system: (i) the collection phase, where we record sensor inputs, and (ii) the configuration phase, where we determine the particle counts of all tasks. During phase (i), we collect input data from the sensors of the actual system, which we replay repeatedly during (ii) in a hardware-in-the-loop simulation. The configuration phase runs on the target system. This paper focuses on the configuration phase (ii).


We present two alternative approaches to automatic configuration based on execution-time fairness (Sec.~\ref{subsec:exec-time-config}) and particle fairness (Sec.~\ref{subsec:particle-config}). In both cases, we seek a particle count $n_\tau$ to use in the inference of each task $\tau$ based on its importance value $v_\tau$, using fair utilization. For simplicity, we assume the user provides a task-to-core mapping $\mathcal{M}$ specifying on which core each task runs (such that $\mathcal{M}_\tau \in \mathcal{U}$). Also, we assume the existence of a function \textsc{RunTasks} that runs the tasks with pre-recorded sensor data to measure the WCET $W_\tau$ of each task $\tau$. Finally, we use a configurable safety margin factor $\delta$ to protect against outliers in the WCET observations.

\subsection{Execution-Time Fairness}
\label{subsec:exec-time-config}

\noindent We perform two steps to achieve execution-time fairness. First, we compute fair execution time budgets $B_\tau$ for each task $\tau$ based on its importance $v_\tau$, such that all tasks remain schedulable. Second, we maximize the particle counts while ensuring the WCET of each task $\tau$ is within its budget.

We define the \textsc{ComputeBudgets} function in Algorithm~\ref{alg:config-exec} (lines~\ref{alg:config-exec:budgets1}-\ref{alg:config-exec:budgets2}) to compute the execution time budgets of tasks with fair utilization. Our approach assumes partitioned scheduling, which means we can examine tasks assigned to different cores independently. For each core $c_i$, we use the sensitivity analysis in the C-space of Bini et al.~\cite{bini2008sensitivity} to compute a maximum change $\lambda_i$ to the execution times of tasks for which they remain schedulable, given a direction vector $\mathbf{d}$ (initially, we assume all execution times are zero). We define $\mathbf{d}$ based on the importance values of the tasks (line~\ref{alg:config-exec:d-vec}) and pass it to the \textsc{SensitivityAnalysis} function to compute $\lambda_i$ (line~\ref{alg:config-exec:sensitivity}). Then, we pick the minimum lambda among all cores (line~\ref{alg:config-exec:lambda}) and use this to get fair execution time budgets for all tasks (line~\ref{alg:config-exec:budgets}).

\begin{figure}[t!]
\begin{algorithm}[H]
\vspace{-0.25em}
\caption{Function to determine the number of particles to use for execution-time fairness.}
\label{alg:config-exec}
\begin{algorithmic}[1]
\footnotesize
\Function{ComputeBudgets}{$\mathcal{M}, \mathcal{T}$}\label{alg:config-exec:budgets1}
    \For{$c_i \in \mathcal{U}$}
        \State $\mathbf{d} \gets \{v_\tau\ |\ \tau \in \mathcal{T} \land \mathcal{M}_\tau = c_i\}$\label{alg:config-exec:d-vec}
        \State $\lambda_i \gets \Call{SensitivityAnalysis}{\mathbf{d}}$\label{alg:config-exec:sensitivity}
    \EndFor
    \State $\lambda \gets \min_i \lambda_i$\label{alg:config-exec:lambda}
    \State $\mathbf{return}\ \{(\tau, v_\tau \cdot \lambda)\ |\ \tau \in \mathcal{T}\}$\label{alg:config-exec:budgets}
\EndFunction\label{alg:config-exec:budgets2}
\Function{Configure$_{EF}$}{$\mathcal{M}, \mathcal{T}$} \label{alg:config:exec1}
    \State $L, U \gets \{(\tau, 1)\ |\ \tau \in \mathcal{T}\}, \{(\tau, \infty)\ |\ \tau \in \mathcal{T}\}$
    \State $N \gets \{(\tau, 1)\ |\ \tau \in \mathcal{T}\}$
    \State $A \gets \{\tau\ |\ \tau \in \mathcal{T} \land P_\tau = \emptyset\}$ \label{alg:config:a-init}
    \State $B \gets \Call{ComputeBudgets}{\mathcal{M}, \mathcal{T}}$
    \State $F \gets \{\tau\ |\ \tau \in \mathcal{T} \land v_\tau = 0\}$ \label{alg:config:f-init}
    \While{$A \neq \emptyset$}
        \State $W \gets \Call{RunTasks}{\mathcal{M}, \mathcal{T}}$ \label{alg:config:run-tasks1}
        \For{$\tau \in A$}
            \If{$L_\tau + 1 < U_\tau$} \label{alg:config:binsearch1}
                \State $L_\tau, U_\tau \gets \begin{cases}
                    L_\tau, N_\tau & \text{if }\frac{W_\tau}{\delta} > B_\tau\\
                    N_\tau, U_\tau & \text{otherwise}
                \end{cases}$
                \State $N_\tau \gets \begin{cases}
                    2 \cdot L_\tau & \text{if }U_\tau = \infty\\
                    \lfloor \frac{L_\tau + U_\tau}{2} \rfloor & \text{otherwise}
                \end{cases}$
            \Else\ 
                $N_\tau, F \gets L_\tau, F \cup \{\tau\}$\label{alg:config:conclusion}
            \EndIf
        \EndFor \label{alg:config:binsearch2}
        \State $A \gets \{\tau\ |\ \tau \in \mathcal{T} \land P_\tau \subseteq F\} \setminus F$ \label{alg:config:a-upd}
    \EndWhile
    \State $\mathbf{return}\ N$
\EndFunction \label{alg:config:exec2}
\end{algorithmic}
\end{algorithm}
\vspace{-2em}
\end{figure}

\begin{figure}[t!]
\begin{algorithm}[H]
\vspace{-0.25em}
\footnotesize
\caption{Computing the number of particles to use for particle fairness.}
\label{alg:config-part}
\begin{algorithmic}[1]
\Function{Schedulable}{$\mathcal{M}, \mathcal{T}, k$} \label{alg:config:sched1}
    \For{$\tau \in \mathcal{T}$} \label{alg:config:part-loop1}
        $n_\tau \gets \lfloor k \cdot \overline{v}_\tau \rfloor$
    \EndFor \label{alg:config:part-loop2}
    \State $W \gets \Call{RunTasks}{\mathcal{M}, \mathcal{T}}$ \label{alg:config:run-tasks2}
    \For{$(\tau, W_\tau) \in W$} \label{alg:config:sched-set-wcet1}
        $C_\tau \gets \frac{W_\tau}{\delta}$
    \EndFor \label{alg:config:sched-set-wcet2}
    \State $\textbf{return}\ \Call{RTA}{\mathcal{M}, \mathcal{T}}$ \label{alg:config:sched-rta}
\EndFunction \label{alg:config:sched2}
\Function{Configure$_{PF}$}
{$\mathcal{M}, \mathcal{T}$} \label{alg:config:particle1}
    \State $L, U \gets 1, \infty$ \label{alg:config:particle-bounds}
    \While{$L + 1 < U$} \label{alg:config:particle-bsearch1}
        \State $k \gets \begin{cases}
            2 \cdot L & \text{if }U = \infty\\
            \lfloor \frac{L + U}{2} \rfloor & \text{otherwise}
        \end{cases}$
        \State $L, U \gets \begin{cases}
            k, U & \text{if }\Call{Schedulable}{\mathcal{M}, \mathcal{T}, k}\\
            L, k & \text{otherwise}
        \end{cases}$
    \EndWhile \label{alg:config:particle-bsearch2}
    \State $\textbf{return}\ \{(\tau, \lfloor L \cdot \overline{v}_\tau \rfloor)\ |\ \tau \in \mathcal{T}\}$
    \label{alg:config:particle-ret}
\EndFunction \label{alg:config:particle2}
\end{algorithmic}
\end{algorithm}
\vspace{-2.5em}
\end{figure}

We use the execution time budget $B_\tau$ as an upper bound for the WCET of the task $\tau$ while trying to maximize its particle count $n_\tau$. We need to carefully adjust the particle counts of tasks, as they may depend on each other. Assume we have a maximum particle count $n_\tau$ for a task $\tau$. Increasing the particle count of a predecessor task $\tau'$ of $\tau$ causes the execution time of $\tau$ to increase. However, if we keep the particle count of all predecessors fixed, the WCET $C_\tau$ of a task $\tau$ is proportional to its particle count $n_\tau$. This enables using binary search to find the maximum $n_\tau$.

We use these observations in the \textsc{Configure}$_{EF}$ function of Algorithm~\ref{alg:config-exec}, which returns the particle count $N_\tau$ to use for each task $\tau$. We keep track of a set of active tasks $A$ with no active predecessors (initialized on line~\ref{alg:config:a-init}). We use $P_\tau$ to denote the set of predecessors of a task $\tau$. Note on line~\ref{alg:config:f-init} that the finished set $F$ is initialized to contain all tasks with importance zero (tasks $\tau$ that perform no inference are assumed to have $v_\tau = 0$). While we have active tasks, we use the \textsc{RunTasks} function to measure the WCETs of all tasks (line~\ref{alg:config:run-tasks1}). We keep track of lower and upper bounds, $L_\tau$ and $U_\tau$, of the particle count for each task $\tau$. For each active task $\tau$, we compare its measured WCET $W_\tau$ to its budget $B_\tau$ and update the bounds accordingly (lines~\ref{alg:config:binsearch1}-\ref{alg:config:binsearch2}). When the binary search concludes, we set the particle count $N_\tau$ to the lower bound and add the task $\tau$ to the set of finished tasks $F$ (line~\ref{alg:config:conclusion}). At the end of each iteration, we update the set of active tasks (line~\ref{alg:config:a-upd}). When the active set $A$ is empty, the resulting particle counts $N$ are returned. Note that the algorithm assumes that the task dependency graph is acyclic.

\subsection{Particle Fairness}
\label{subsec:particle-config}

\noindent To achieve particle fairness, we seek a multiple $k$ such that $n_\tau = \lfloor k \cdot \overline{v}_\tau \rfloor$ for all tasks $\tau$, where $\overline{v}_\tau$ is the normalized importance value 
. When we increase the value of $k$, the execution time of all tasks increases, meaning we can binary search to find the maximum $k$. For a given $k$, we need to determine whether the tasks are schedulable. We do this in the \textsc{Schedulable} function on lines~\ref{alg:config:sched1}-\ref{alg:config:sched2} of Algorithm~\ref{alg:config-part}. We set the number of particles for each task (line~\ref{alg:config:part-loop1}) and run them (line~\ref{alg:config:run-tasks2}). Then, we set the WCET $C_\tau$ of the tasks based on their measured WCETs (line~\ref{alg:config:sched-set-wcet1}) and perform a response time analysis~\cite{bini2008sensitivity} to ensure the tasks are schedulable (line~\ref{alg:config:sched-rta}).

We use the \textsc{Schedulable} function in the particle fairness function \textsc{Configure}$_{PF}$ (lines~\ref{alg:config:particle1}-\ref{alg:config:particle2}). We define the lower and upper bounds $L$ and $U$ on line~\ref{alg:config:particle-bounds}, and we binary search in this interval on lines~\ref{alg:config:particle-bsearch1}-\ref{alg:config:particle-bsearch2}. After the loop, the maximum multiple for which the system is schedulable is stored in $L$. Based on this, we compute the particle count of each $\tau$ on line~\ref{alg:config:particle-ret}.

\section{Automotive Case Study}
\label{sec:case-study}

\noindent To demonstrate the applicability of ProbTime to program time-critical embedded systems, we develop a case study for a localization pipeline in an automotive testbed.
This section first introduces the testbed, followed by an evaluation of ProbTime.

\subsection{Automotive Testbed}

\noindent The automotive industry is undergoing a paradigm shift to face the challenges that emerge through the advent of autonomous driving~\cite{Zefowski}.
These challenges require higher computational capacities and larger communication bandwidth than traditional automotive systems.
This transition leads to a software-defined vehicle where almost all functions of the car will be enabled by software~\cite{Windpassinger:2022}.
To demonstrate the applicability of our proposed methods in this context, we have developed an automotive testbed.

\subsubsection{Hardware}

\begin{figure}[t!]
\centering
\includegraphics[width=0.8\columnwidth]{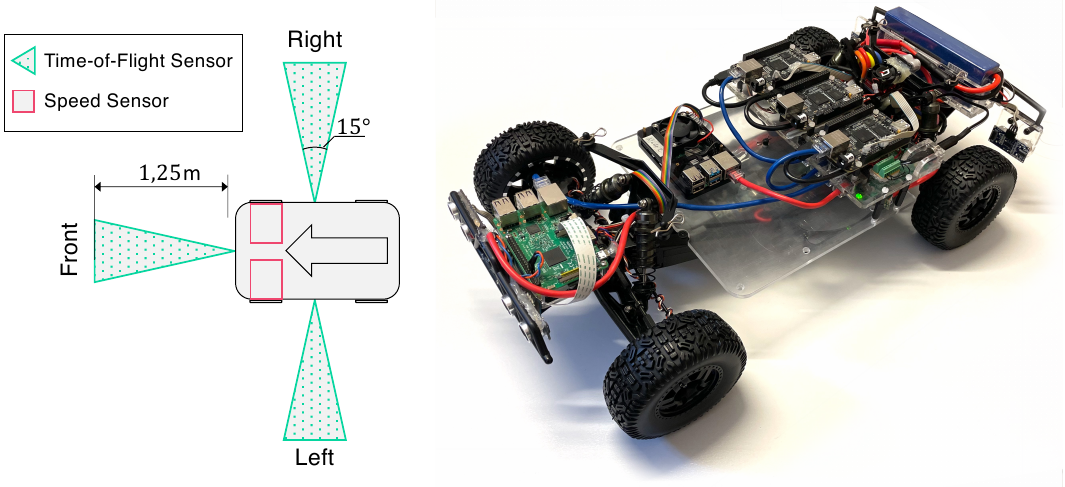}
\caption{Overview of the sensors on the car (left) and a picture of the RC-Car testbed (right).}
\label{fig:CScarPicture}
\vspace{-1em}
\end{figure}

The testbed is designed around a 1:10 scale RC car.
The E/E architecture consists of several distributed compute nodes of different characteristics connected via bus- and network-based interconnects (see Fig.~\ref{fig:CScarPicture}).

Two microcontroller-based compute nodes act as the interface to most actuators and sensors of the car.
High-performance compute nodes provide the required performance to host computation required for Advanced Driving Assistance (ADAS) or Autonomous Driving (AD) workloads.
We use a Raspberry Pi 4B (4 Cortex-A72 cores at 1.5GHz).
These compute nodes are connected by a local Ethernet network.

We use three Time-of-Flight (TOF) sensors installed
on the car, as indicated in Fig.~\ref{fig:CScarPicture}. These sensors have high accuracy but can only measure distances up to $1.25~\text{m}$. The front wheel sensors report the current speed of the car.

\subsubsection{Software}

\begin{figure}[t!]
\centering
\includegraphics[width=0.8\columnwidth]{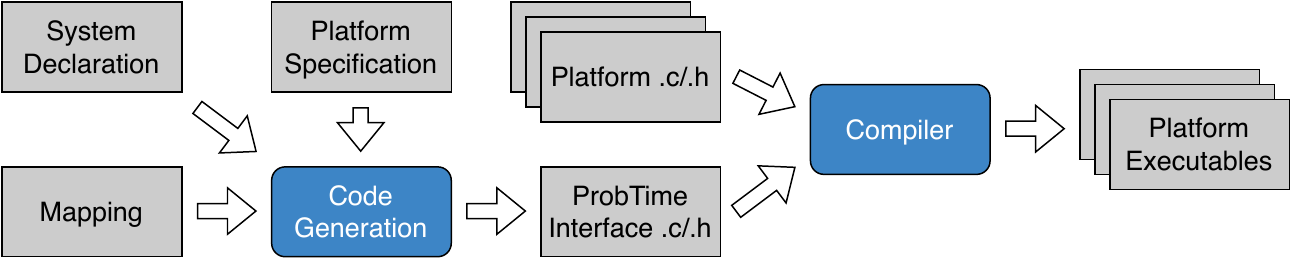}
\caption{Code generation and compilation of the platform code, including a ProbTime interface.}
\label{fig:platformComp}
\vspace{-1em}
\end{figure}

The operating system on the Raspberry Pi 4B compute node is Linux-based with kernel 5.15.55-rt48-v8+ SMP PREEMPT\_RT. The platform software is responsible for reading sensor data and controlling actuators.
This is realized by dedicated periodic tasks distributed on different compute nodes.
Each task is assigned a static priority and is mapped to a CPU core.
Platform software realizes communication between ProbTime tasks by communication buffers located in shared memory.
Additionally, sensor data is written to ProbTime communication buffers and from buffers to actuators. 


We use code generation to produce the files that constitute the interface to a specific ProbTime program (see Fig.~\ref{fig:platformComp}).
Input to the code generation is the system declaration describing the ProbTime program.
The remaining platform software is not affected by code generation.
The compiler produces an executable for each compute node in the final step.




\subsection{Evaluation Questions}
\label{subsec:evaluation-questions}

\noindent To evaluate ProbTime and our automatic configuration, we pose the following evaluation questions:
\begin{enumerate}
\item[Q1] To what extent can ProbTime's timing and probabilistic constructs be used to implement a realistic application?
\item[Q2] What is the trade-off between the number of particles, execution time, and inference accuracy?
\item[Q3] How does execution-time fairness compare to particle fairness concerning allocation of particles and inference time when varying the importance of tasks?
\item[Q4] To what extent and accuracy can the case study---when executed on the physical car---determine the position and avoid collision?
\end{enumerate}

\subsection{Positioning and Braking in ProbTime (Q1)}
\label{subsec:pos-brake}

\noindent We implement a positioning and braking system in ProbTime as a case study. The system estimates the position of the RC car while it is moving using sensor inputs. Further, the system reacts by braking to prevent collisions with obstacles.

\begin{figure}[t!]
\centering
\includegraphics[width=.8\columnwidth]{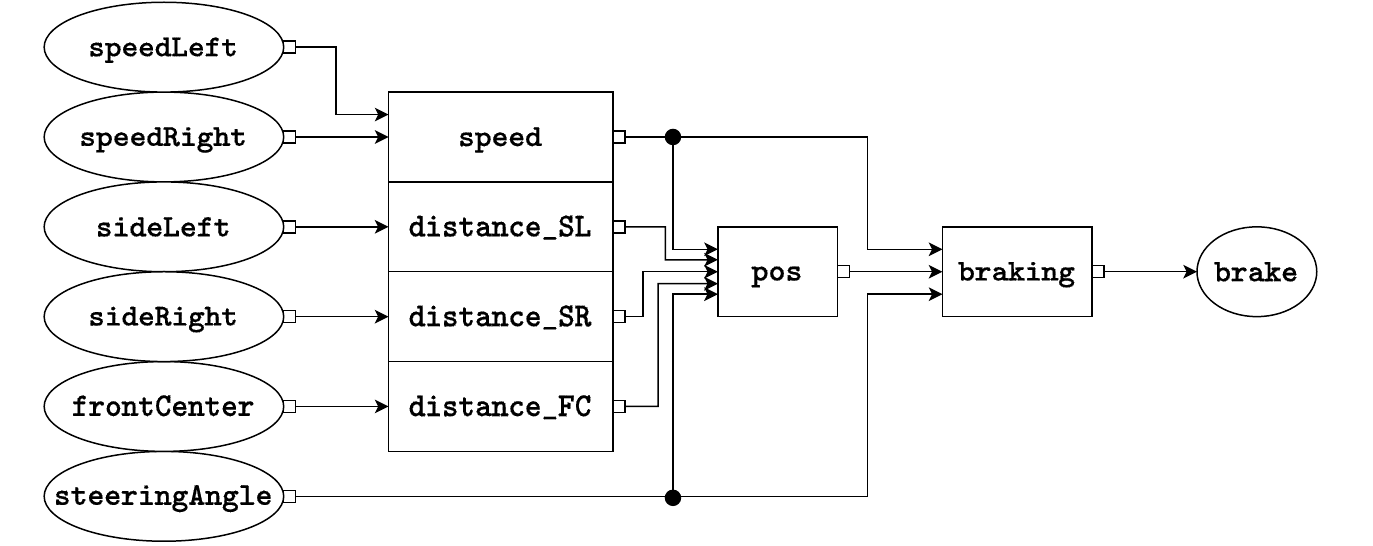}
\caption{A graphical representation of the positioning and braking system.}
\label{fig:posmodel}
\vspace{-1em}
\end{figure}

We present an overview of our positioning and braking system in Fig.~\ref{fig:posmodel}.
The figure shows the sensor inputs on the left-hand side. The steering angle is an input signal from the controller, while the other five sensors are directly related to sensors on the car (as shown in Fig.~\ref{fig:CScarPicture}). We use the \texttt{brake} actuator (right-hand side) to activate the emergency brakes.


The squares of Fig.~\ref{fig:posmodel} represent the tasks. 
We instantiate all distance estimation tasks (\verb|distance_SL|, \verb|distance_SR|, and \verb|distance_FC|) from the same template. All tasks have a period of \SI{500}{\milli\second} except \texttt{pos}, which has a period of \SI{1}{\second}.


We use the \texttt{speed} task to estimate the car's average speed since the last estimation based on speed observations from the wheels.
Preliminary experiments show that the speed sensors are inaccurate at low speeds and that the top speed is reached almost instantaneously when accelerating. Due to the predictable behavior of the speed, we model it as being in one of three states: stationary, max speed, or transitioning (either accelerating or decelerating). This is simpler than using Bayesian linear regression (as we did in the speed model in Listing~\ref{lst:train-speed-model}) and computationally cheaper.

Our positioning task estimates the x- and y-coordinates of the car's center. It also estimates the direction the car faces relative to an encoded map of the environment (provided by the user). The model starts from a prior belief of the position (initially, a known starting position) and estimates a trajectory along which the car moved until the current logical time of the task based on steering angle observations and speed estimations. We update our belief by comparing each input distance estimation with what the sensor would have observed at that timestamp, given that the car followed the estimated trajectory. We implement this using Bayesian linear regression.

The \texttt{braking} task uses a probabilistic model to estimate the distance until a collision occurs, given steering angle observations, an encoded map, and the speed and position estimations. We activate the emergency brakes if the median distance is estimated to be below a fixed threshold. 

In response to Q1, the current case study demonstrates that a non-trivial application can be efficiently implemented in ProbTime (using 600 lines of code).

\subsection{Experiments and Results}
\label{subsec:experiments-results}

\noindent To answer evaluation questions Q2, Q3, and Q4, we perform three experiments on the positioning system of Fig.~\ref{fig:posmodel}. We want to run as many particles per iteration as possible for the first experiment. Therefore, we run this on an Intel(R) Xeon(R) Gold 6148 CPU with 64 GB RAM using Ubuntu 22.04. We use the Raspberry Pi on the car for the second and third experiments. The first two experiments use pre-recorded data in a simulation, while the third uses live data. When discussing results or settings, we only consider the \texttt{pos} and \texttt{braking} tasks, as the other tasks perform no inference. The \texttt{pos} and \texttt{braking} tasks run on separate cores.

\subsubsection{Positioning Inference Accuracy (Q2)}

\noindent We investigate how the number of particles impacts the inference accuracy and execution times. To be able to run more particles per inference, we perform a \emph{slow mode} simulation where the replaying of observation data and the ProbTime tasks consider time to pass slower relative to the wall time. We use position estimates as a measure of inference accuracy because they are easily compared to a known final position from pre-recorded data. In this experiment, we vary the particle count of the \texttt{pos} task while fixing the particle count of \texttt{braking}.

\begin{figure}[t!]
\centering
\includegraphics[width=0.8\columnwidth]{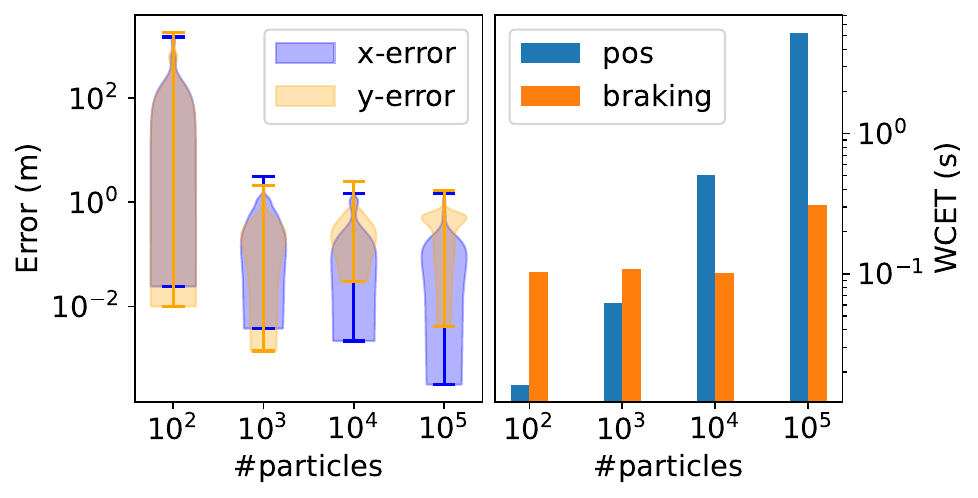}
\caption{Results showing the estimation error along the x- and y-axis of the car (left) and the WCETs (right) over ten runs, varying the number of particles in the positioning model. Both figures have a logarithmic x- and y-axis.}
\label{fig:convergence}
\vspace{-1em}
\end{figure}

In Fig.~\ref{fig:convergence}, we present the average position error along the x- and y-axis (left) and the WCETs (right) for the two tasks over $100$ runs with varying particle counts in the positioning task. Note that the positioning model often fails to track the car when using $10^2$ particles. The estimation error along both axes decreases drastically as we increase the number of particles, mainly when increasing from $10^2$ to $10^3$ particles. The increased variance going from $10^4$ to $10^5$ particles is likely due to inaccuracies in our model. The WCET of the \texttt{pos} task scales linearly with the number of particles (note the logarithmic axes), while the \texttt{braking} task becomes noticeably slower when we increase the number of particles to $10^5$. This is because it processes a position distribution sent from the \texttt{pos} task (i.e., \texttt{braking} depends on \texttt{pos}).

In connection to Q2, we have seen the impact of particle count on the inference accuracy. Few particles lead to a big uncertainty, and the benefit of increasing the particle count is negligible past a certain point. We also see that the number of particles is linearly connected to the WCET and that the particle count of a task may impact the WCET of other tasks.

\subsubsection{Automatic Configuration (Q3)}

\noindent To compare the automatic configuration using execution-time and particle fairness, we vary the ratio of importance between the \texttt{pos} and \texttt{braking} tasks. For each ratio, we measure the particle count and WCET of the configured tasks and the number of iterations needed to configure the system.

\begin{figure}[t!]
\centering
\includegraphics[width=0.8\columnwidth]{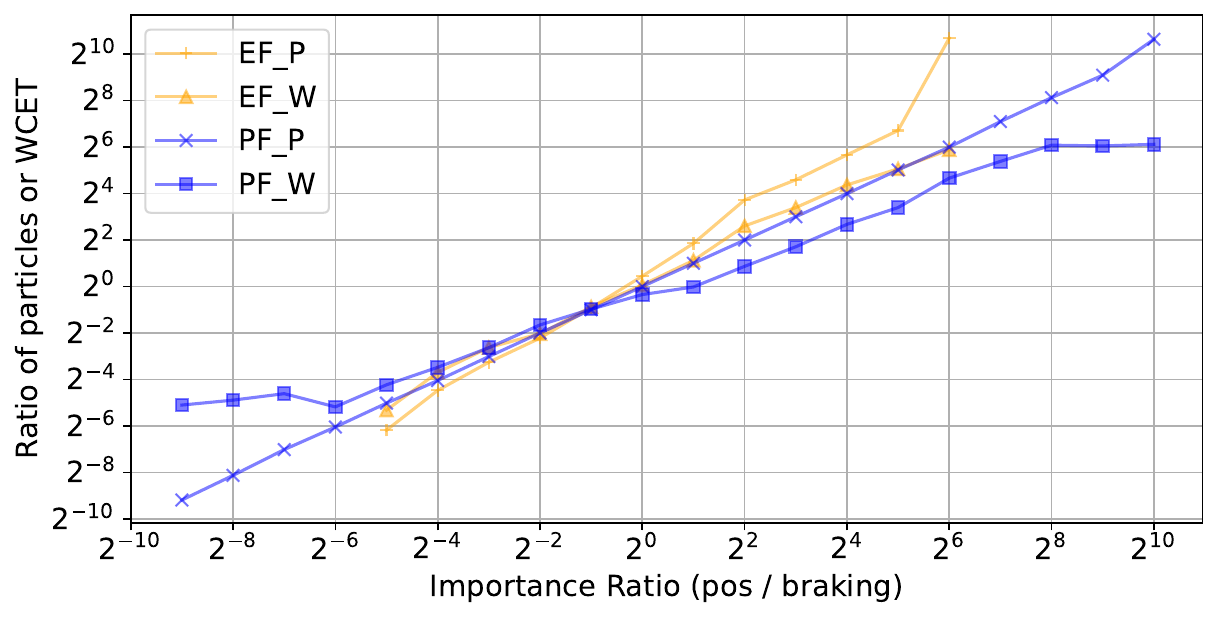}
\caption{The ratio of particles (using suffix \texttt{P}) and worst-case execution time (suffix \texttt{W}) when varying the importance ratio. We use prefix \texttt{EF} for execution-time fairness results and \texttt{PF} for particle fairness. Note the logarithmic x- and y-axes.}
\label{fig:auto-config}
\vspace{-1em}
\end{figure}

The resulting ratio of particle count and execution time is presented in Fig.~\ref{fig:auto-config}.
In our experiment, we vary the importance ratio between tasks by powers of two until the starting state cannot be scheduled. As expected, both approaches to fairness have one property scaling at the same rate as the importance ratio (WCET ratio for execution-time fairness and particle count ratio for particle fairness).

The WCET ratio for particle fairness peaks when the ratio imbalance is too large (e.g., when the ratio is $2^8$) as the more important task fully utilizes its core. Increasing the ratio imbalance past this point only causes the less important task to run fewer particles. For execution-time fairness, this imbalance results in the less important task being allocated an insufficient budget due to fair utilization.

Execution-time fairness needs $48.8 \pm 11.8$ iterations (mean and standard deviation) to configure all tasks compared to $16.5 \pm 2.8$ for particle fairness. Recall from Sec.~\ref{sec:configuration} that execution-time fairness considers task dependencies while particle fairness runs independently of them.


For our model, the problem of task dependencies is hardly noticeable. However, what if the \texttt{speed} task also performed inference? To investigate this, we configure a variant of the positioning and braking model where \texttt{speed} performs inference, with it running on a separate core from \texttt{pos} and \texttt{braking}, and all three tasks have the same importance. We find that the tasks all get $385$ particles using particle fairness, whereas using execution-time fairness, \texttt{pos} gets $21$ particles, \texttt{braking} gets $95$ particles, and \texttt{speed} gets $24199$ particles. For execution-time fairness, the \texttt{speed} task has time to run many particles, which slows down the other tasks that depend on its output. This issue could be mitigated by setting an upper bound on the particle count (machine-specific) or by giving \texttt{pos} and \texttt{braking} higher importance.

\begin{figure}[t!]
\centering
\includegraphics[width=0.8\columnwidth]{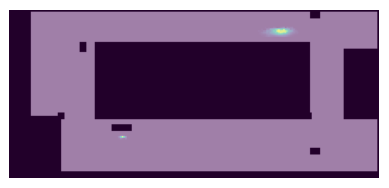}
\caption{The encoded map (\SI{18.6}{\meter} x \SI{9.5}{\meter}) of the corridor in which we drive the car.
The overlay shows particles spreading when less information is available (top-right) and gathering when close to distinguishable obstacles (bottom-left).}
\label{fig:corridor}
\vspace{-1em}
\end{figure}

In connection to Q3, we argue that particle fairness is preferable. First, the problems above only apply to execution-time fairness. Second, there is a direct connection between particle count and inference accuracy. Therefore, we argue that reasoning about the allocation of particles directly seems more natural than doing it in terms of execution time.

\subsubsection{Positioning and Braking (Q4)}

\noindent For this experiment, we run the tasks on the physical car. To evaluate the positioning model, we drive the car around in a corridor (Fig.~\ref{fig:corridor}).

As we measure the car's position by hand, we can only make measurements when the car is stationary. Therefore, we drive it from a known starting position and compare the actual final position to the last position estimated by the model. We drive the car in a clockwise direction for a full lap ten times using two distinct scenarios to illustrate how environmental details impact the accuracy of the position estimates. In scenario 1, we start at the bottom-left corner and drive past the starting point. In scenario 2, we begin in the top-right corner and stop along the long corridor at the top. During preliminary evaluations, we found that using importance $4$ for \texttt{pos} and $1$ for \texttt{braking} yields good results. The configured system runs inference using $1312$ particles in \texttt{pos} and $328$ in \texttt{braking}.

We evaluate the automated braking by driving the car around in the corridor. We find that the emergency brakes are conservative---while they prevent collisions, they often activate in situations where it is unnecessary. This is due to uncertainty in the position estimations. Due to the conservative nature of the \texttt{braking} task, we turn off the \texttt{brake} actuator when evaluating the positioning model.

We present the positioning results in Table~\ref{table:pos-results}. The error along the y-axis is small in scenario 2 because both side sensors are in range in the narrow corridor where the car stops, while the x-axis error is large because the front distance sensor is outside of its range in the corridor. Despite the short range of our sensors, the lack of distinguishing details in the environment, and the limited computational power of the car testbed, our model successfully tracks the car for a full lap in the corridor. Tracking the car in itself is not a novelty, but it shows that the methodology works in practice for a non-trivial application.

\begin{table}[t!]
\centering
\caption{Results showing the average error in the x- and y-direction (in meters) and the Euclidean distance for our two scenarios, with standard deviations.}
\label{table:pos-results}
\begin{tabular}{c|ccc}
Scenario & x-axis error & y-axis error & Euclidean distance\\
\hline
1  & $0.50 \pm 0.58$ & $0.28 \pm 0.23$ & $0.63 \pm 0.56$\\
2  & $1.19 \pm 0.79$ & $0.24 \pm 0.14$ & $1.21 \pm 0.80$
\end{tabular}
\vspace{-1em}
\end{table}


\section{Related Work}

\noindent Although surprisingly little has been done in the intersection of probabilistic programming and languages for real-time systems, there is a large body of work in each category. This section gives a summary of the closest related work.



\emph{Probabilistic programming languages.} Languages for performing Bayesian inference date back to the early 1990s, where BUGS~\cite{GilksEtAl:1994} is one of the earliest languages for Bayesian networks. The term probabilistic programming was coined much later but is now generally used for all languages performing Bayesian inference. Starting with the Church language~\cite{GoodmanEtAl:2008}, a new, more general form of probabilistic programming languages emerged, called \emph{universal PPLs}. In such languages, the probabilistic model is encoded in a Turing complete language, which enables stochastic branching and possibly an infinite number of random variables. There exist many experimental PPLs in this category, e.g., Gen~\cite{CusumanoTownerEtAl:2019}, Anglican~\cite{TolpinEtAl:2016}, Turing~\cite{GeXuGhahramani:2018}, WebPPL~\cite{GoodmanStuhlmuller:WebPPL}, Pyro~\cite{BinghamEtAl:2019}, and CorePPL~\cite{LundenEtAl:2022:RootPPL}. Other state-of-the-art PPL environments are, e.g., Stan~\cite{CarpenterEtAl:2017} and Infer.NET. Similar to many of these environments, ProbTime also makes use of the standard core constructs (\verb|sample|, \verb|observe|, and \verb|infer|), where the key difference is that time and timeliness are also taken into consideration in ProbTime.

\emph{Real-time programming languages.} There exists a quite large body of literature on programming languages that incorporate time and timing as part of the language semantics. Ada~\cite{BurnsWellings:2007} has explicit support for timed language constructs, and the real-time extension to Java~\cite{bollella2000real} supports time and event handling.
Recently, Timed C was proposed in the research literature to incorporate timing constructs directly in the language. The Timed C work was later extended to support sensitivity analysis~\cite{NatarajanEtAl:2019}. Our work has been inspired by Timed C; the semantics for \verb|periodic| is loosely based on \verb|sdelay| in Timed C, although the implementation is different. As in Timed C, ProbTime introduces tagging with timestamps.
%
There are several other languages and environments that have timestamped and tagged values as central semantic concepts. For instance, PTIDES~\cite{EidsonEtAl:2011} extends the discrete-event (DE) model.
Lingua Franca~\cite{LohstrohEtAl:2021} is another recent effort that introduces deterministic actors called reactors, whose only means of communication is through an ordered set of reactions.

\emph{Synchronous programming.} There are also other approaches to programming real-time systems where time is abstracted as ticks. Specifically, the synchronous programming paradigm~\cite{BenvenisteBerry:1991} implements such an approach, where ProbZelus~\cite{BaudartEtAl:2020} is a recent programming language extending synchronous reactive programming with probabilistic constructs. The ProbZelus work is based on the delayed sampling concept~\cite{LawrenceEtAl:2018} to automatically make inference efficient when conjugate priors can be exposed in the model.
A key difference between ProbZelus and ProbTime is our concept of fairness, where the number of particles used in inference is automatically adjusted based on user-provided importance values.

\emph{Reward-based scheduling.} Our automatic configuration is related to reward-based scheduling~\cite{aydin2001optimal}, where tasks consist of a mandatory and an optional part. A non-decreasing reward function maps the time spent running the optional part to a real value. This approach aims to find a schedule for maximizing the reward while all tasks remain schedulable. In ProbTime, inference is the optional part of tasks as we can vary how many particles we use. The importance values assigned to tasks implicitly encode a reward function.

\section{Conclusion}

\noindent In this paper, we introduce a new kind of language called \emph{real-time probabilistic programming language (RTPPL)}. To demonstrate this new kind of language, we develop and analyze a domain-specific RTPPL called ProbTime. We also introduce the concepts of fairness in RTPPLs and illustrate the strength of particle fairness.

%

\iftoggle{anonymous}{}{
\section*{Acknowledgment}

\noindent We would like to thank John Wikman, Linnea Stjerna, Gizem Çaylak, Viktor Palmkvist, Anders Ågren Thuné, Thilanka Thilakasiri, and Rodolfo Jordão for their valuable feedback.

This project is financially supported by the Swedish Foundation for Strategic Research (FFL15-0032 and RIT15-0012). The research has also been carried out as part of the Vinnova Competence Center for Trustworthy Edge Computing Systems and Applications (TECoSA) at the KTH Royal Institute of Technology.
}

\bibliography{IEEEabrv,references,dbro}

\end{document}